\documentclass[ footinbib,
                twocolumn,
                showpacs,
                aip,apl,reprint,
                superscriptaddress,
                floatfix,
                raggedbottom,
                citeautoscript]{revtex4-1}

\usepackage{graphicx} \usepackage{amsmath, amssymb, bm}
\usepackage{dcolumn} \usepackage{color}
\usepackage{eso-pic}
\usepackage{fix-cm}

\usepackage[capbesideposition=right]{floatrow}
\usepackage{multirow}
\usepackage{sidecap}

\usepackage{marginnote}
\usepackage{ulem} 

\newcommand{\sss}{\scriptscriptstyle}
\newcommand{\sst}{\scriptstyle}

\newcommand{\stext}[1]{\sss \text{#1} \sst}

\newcommand{\veps}{\varepsilon}

\makeatother

\begin{document}

\title{Structural and optical properties of cobalt slanted columnar thin films\\ conformally coated with graphene by chemical vapor deposition}

\author{Peter M.~Wilson}
\affiliation{Department of Chemistry, and Center for Nanohybrid Functional Materials,\\ University of Nebraska-Lincoln, Lincoln, Nebraska, 68588, USA}
\author{Daniel~Schmidt}
\affiliation{Department of Electrical Engineering, and Center for Nanohybrid Functional Materials, University of Nebraska-Lincoln, Lincoln, Nebraska, 68588, USA}
\author{Eva~Schubert}
\affiliation{Department of Electrical Engineering, and Center for Nanohybrid Functional Materials, University of Nebraska-Lincoln, Lincoln, Nebraska, 68588, USA}
\author{Mathias~Schubert}
\affiliation{Department of Electrical Engineering, and Center for Nanohybrid Functional Materials, University of Nebraska-Lincoln, Lincoln, Nebraska, 68588, USA}
\author{Alexander~Sinitskii}
\affiliation{Department of Chemistry, and Center for Nanohybrid Functional Materials,\\ University of Nebraska-Lincoln, Lincoln, Nebraska, 68588, USA}
\author{Tino~Hofmann}
\email{thofmann@engr.unl.edu}
\affiliation{Department of Electrical Engineering, and Center for Nanohybrid Functional Materials, University of Nebraska-Lincoln, Lincoln, Nebraska, 68588, USA} \homepage{http://ellipsometry.unl.edu}

\date{\today}

\begin{abstract}
A slanted cobalt sculptured columnar thin film was fabricated using glancing angle deposition, and coated subsequently with graphene using a low temperature chemical vapor deposition process. The graphene deposition process preserves shape and geometry of the sculptured thin film, which was confirmed by scanning electron microscopy. According to the Raman spectroscopy results, the graphene coating is two to three monolayers thick and has a high defect concentration. The graphene coverage within the sculptured thin film is determined from generalized spectroscopic ellipsometry using a generalized anisotropic Bruggeman effective medium approximation. The graphene coverage as well as structural parameters of the thin film agree excellently with electron microscopy and Raman observations, and suggest that the graphene coating is conformal.
\end{abstract}


\maketitle

Metallic sculptured thin films (STFs) provide a versatile platform for highly sensitive optical sensors based on birefringence changes upon analyte adsorption within the nanostructures \cite{RodenhausenOE20_2012, KasputisJCC_2013}. The functionalization and stabilization of such three-dimensional (3D) nanostructured surfaces using conformal surface coatings offers interesting practical applications.
Graphene, a two-dimensional (2D) carbon allotrope, that is known for its excellent mechanical properties\cite{LeeS321_2008,BunchS315_2007} and, for example, its ability to form protective corrosion-inhibiting  barriers on metals\cite{RuoffACSN5_2011,BolotinACSN6_2012} would be an excellent candidate if a coating of the 3D nanostructured surface can be achieved. In addition, a graphene coating on STFs would provide a novel avenue for chemical functionalization, for example, via diazonium chemistry,\cite{FarmerNL9_2009,SinitskiiACSN4_2010} which could be used to increase the selectivity of the STFs to analytes of interest.
The investigation of graphene deposited on metallic STFs therefore is of high interest and no reports describing the optical and structural properties of graphene coated STFs exist in the literature so far.

\begin{figure}
\centering
\includegraphics[keepaspectratio=true,width=8.502cm, clip, trim=0 0 0 0 ]{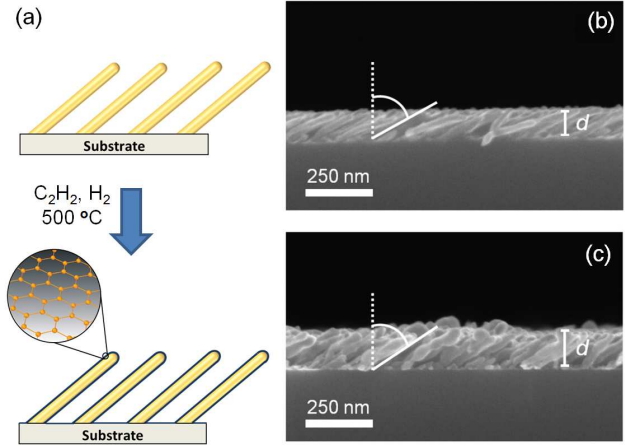}
\caption{A schematic of the conformal graphene coating is shown in (a). High-resolution cross-section SEM micrographs obtained from the Co SCTF sample before (b) and after (c) the CVD process demonstrate that the 3D nanostructured geometry of the film remains intact after the CVD process. The best-model thickness for the film obtained by GSE is indicated as $d$ (see also Tab.~\ref{tab:results}). During the CVD growth the slanting angle slightly decreases from 57.6$^{\circ}$ to 49.2$^{\circ}$ while the nanorod length remains constant.\vspace{-0.4cm}}
\label{fig:SEM}
\end{figure}

In this letter, we demonstrate the fabrication of highly spatially coherent 3D metal nanostructures coated with multilayer graphene using a chemical vapor deposition (CVD) process. The quality of the graphene coating is inspected using confocal Raman microscopy while the optical and structural properties of the nanostructured thin film are investigated using scanning electron microscopy (SEM) and generalized spectroscopic ellipsometry (GSE) \cite{SchmidtOL34_2009,SchmidtJAP105_2009,SchmidtAPL94_2009,SchmidtAPL100_2012,SchmidtJAP114_2013}. It is demonstrated that an augmented anisotropic Bruggeman effective medium approximation (AB-EMA) provides an accurate description of the anisotropic optical response of the STF, which changes its birefringence behavior from biaxial to uniaxial upon graphene deposition. The AB-EMA further allows for the determination of the graphene coverage across the 3D STF surface. The structural parameters obtained from SEM, Raman and through the AB-EMA-based analysis are in excellent agreement. We conclude that the CVD process leads to a conformal graphene coating of the 3D structures with thickness corresponding to 2-3 graphene monolayers.

Cobalt slanted columnar thin films (SCTFs) were fabricated by electron-beam glancing angle deposition (GLAD) in a custom-built ultra-high vacuum chamber at room temperature. The nanocolumnar structures were deposited onto a low-doped $n$-type (001) silicon substrate at a deposition angle of 85$^{\circ}$ \cite{SchmidtJAP105_2009}. Immediately after GLAD growth the sample was transferred to a custom-built CVD system \cite{WilsonJMCC2013}. For the CVD process, acetylene was used as the hydrocarbon precursor due to its ability to decompose to graphene at low temperatures, which is shown schematically in Fig.~\ref{fig:SEM}(a). During the deposition the temperature was raised to 350~$^{\circ}$C under 3.4~mTorr of hydrogen, then 3.0~mTorr of acetylene was added to the hydrogen and the furnace was raised to 500 $^{\circ}$C for 1~min.
After the CVD process the sample was characterized using angle-resolved GSE,\cite{SchmidtOL34_2009} Raman, and SEM.
GSE measurements were carried out in the spectral range from $\lambda=$ 400 to 1650~nm using a commercial instrument equipped with a automated sample rotation stage (M2000VI, J.A.~Woollam Co.~Inc.). Mueller matrix data were obtained for four different angles of incidence $\Phi_{\stext{a}}=45^{\circ},55^{\circ}, 65^{\circ},$ and $75^{\circ}$ for a  complete in-plane sample rotation from $0^{\circ}\le\varphi\le360^{\circ}$ in steps of 6$^{\circ}$ \cite{SchmidtOL34_2009}. In addition to the GSE measurements, unpolarized micro Raman scattering experiments were performed using a commercial Raman microscope (Thermo Scientific DXR) with a 532~nm laser and a 1~$\mu$m spatial resolution in order to assess the quality and morphology of the graphene grown on the 3D nanostructures. High-resolution cross-section SEM micrographs were obtained using a Hitachi S4700 field emission SEM.

\begin{figure}
\includegraphics[keepaspectratio=true,width=7.01cm, clip, trim=0 0 0 0 ]{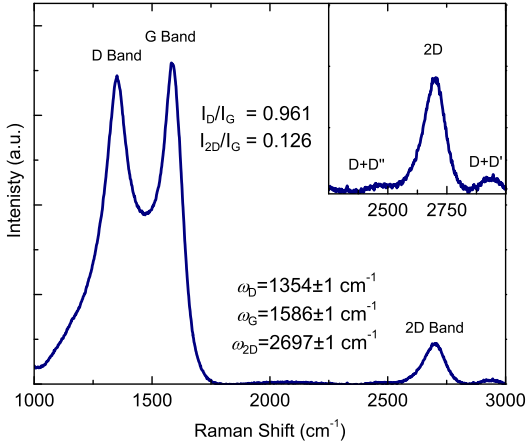}
\caption{Experimental Raman spectra (dotted lines) obtained from the Co SCTF after the graphene CVD (see Fig.~\ref{fig:SEM}(c)). Three distinct Raman bands are identified as D-, G-, and 2D-band. In the vicinity of the 2D resonance two subtle features can be observed, which are identified as D+D$^{\prime\prime}$ and D+D$^{\prime}$ resonances (inset).\vspace{-0.4cm}
}   \label{fig:raman}
\end{figure}

Figure~\ref{fig:SEM} depicts the cross-section SEM micrographs of the Co SCTF sample before (b) and after (c) the graphene CVD.
The 3D nanostructure geometry remains intact after the graphene CVD process. As can be inferred from Figs.~\ref{fig:SEM} (b) and (c), however, a slight decrease of the slanting angle from 57.6$^{\circ}$ to 49.2$^{\circ}$ occurs. Note that control experiments carried out without introducing the hydrocarbon precursor resulted in the disintegration of the nanostructures. The slanting angle change is tentatively associated with the acetylene decomposition on the Co surface, in particular at the interface of the nanocolumns and the Si substrates.

The Raman data obtained after graphene deposition are presented in Fig.~\ref{fig:raman}. Three distinct Raman bands identified as D-, G-, and 2D- bands can be observed \cite{FerrariNN8_2013}. More subtle features are present in the vicinity of the second-order D-band (inset Fig.~\ref{fig:raman}), which are due to D+D$^{\prime}$ and D+D$^{\prime\prime}$ phonon  excitations \cite{FerrariNN8_2013}.
The Raman spectrum for the graphene coated Co SCTFs resembles neither typical monolayer nor multilayer graphene Raman data nor does it show the fingerprints of graphite\cite{FerrariPRL97_2006, FerrariNN8_2013}.
The presence of carbon nanotubes can be ruled out based on the fact that resonances typical for carbon nanotube Raman spectra, e.g., the splitting of the G-band does not occur here\cite{FerrariSSC143_2007}.
The resonance energies of the D-, G-, and 2D- bands and their intensity ratios are given in Fig.~\ref{fig:raman}.

\begin{figure*}
\fcapside[13.5cm]
{\caption{Experimental (circles) and best-match calculated (solid lines) GSE data of the as-grown GLAD Co SCTF (left; Fig.~\ref{fig:SEM} (b)) and the same SCTF after graphene CVD (right; Fig.~\ref{fig:SEM} (c)) as a function of the sample azimuth angle $\varphi$ shown exemplarily for $\lambda$ = 630~nm at three different angles of incidence $\Phi_{\stext{a}}$ = 55$^{\circ}$, 65$^{\circ}$, and 75$^{\circ}$. The elements $M_{14}$ and $M_{24}$ are magnified indicated by the numbers in the respective panel.}\label{fig:MM}}
{\includegraphics[keepaspectratio=true,width=6.601cm, clip, trim=0 0 0 0 ]{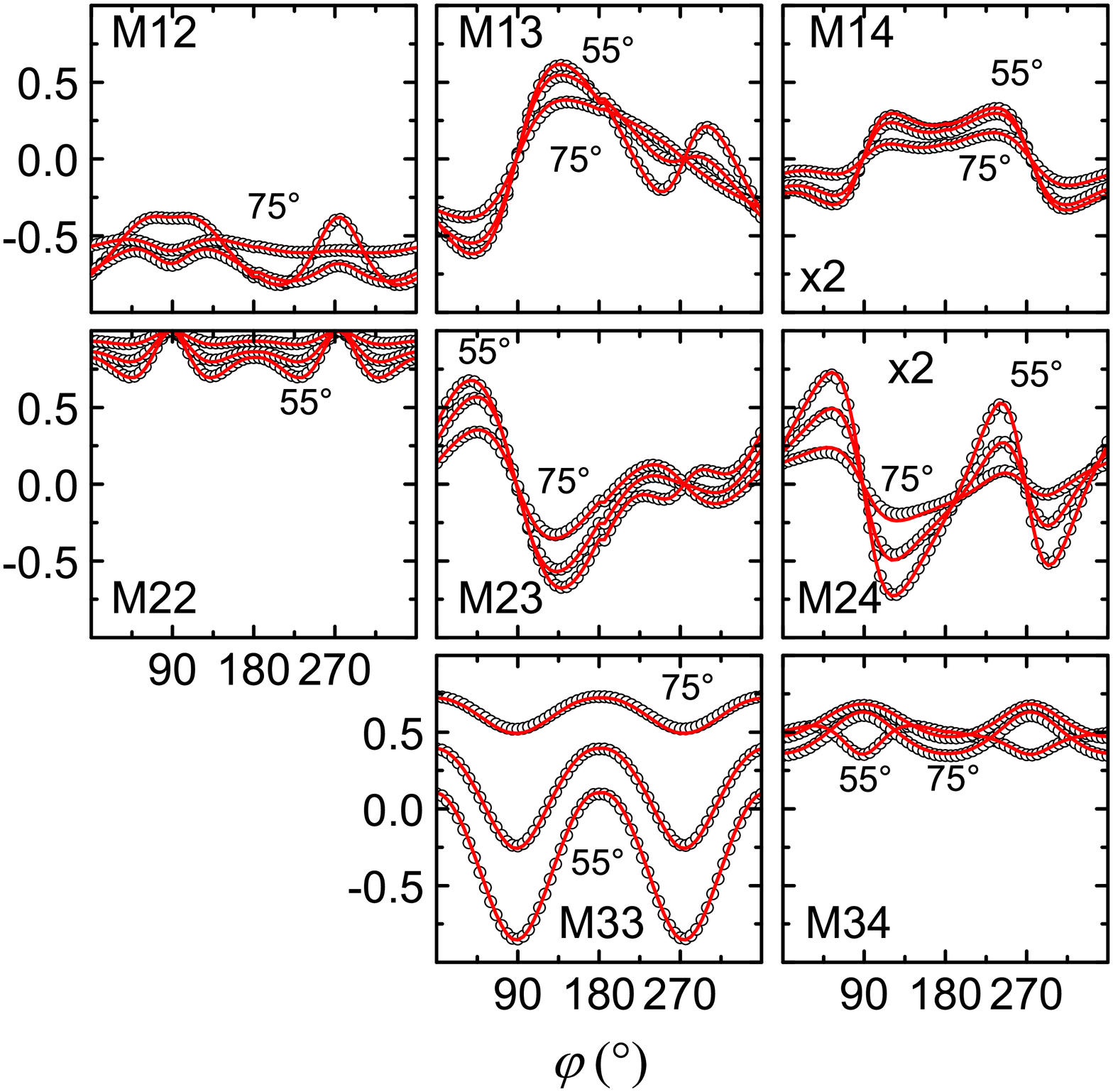}\hspace{0.2cm}
\includegraphics[keepaspectratio=true,width=6.601cm, clip, trim=0 0 0 0 ]{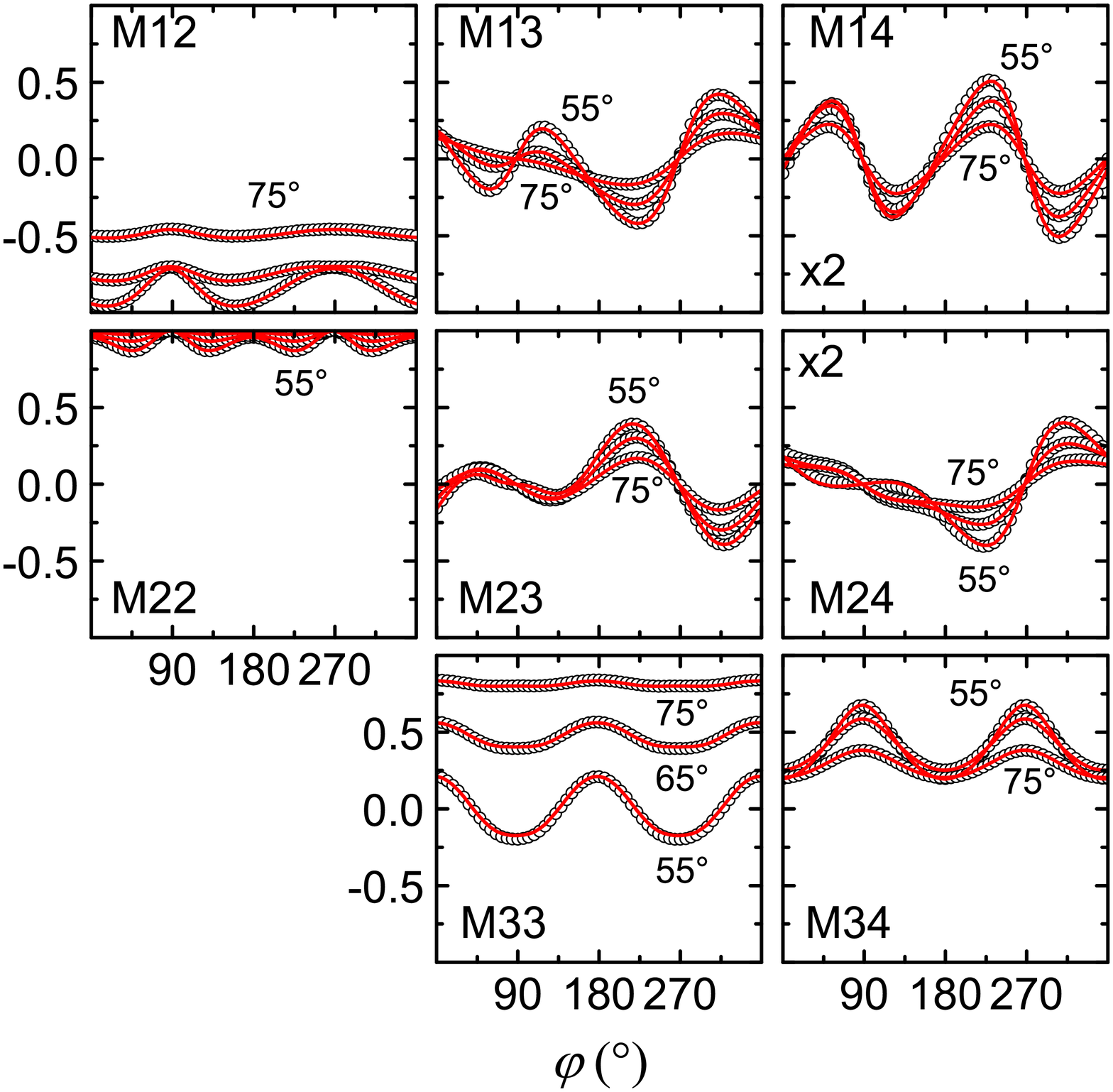}}\vspace{-0.3cm}
\end{figure*}

The comparatively strong Raman D-band ($I_{\stext{D}}/I_{\stext{G}}$=0.921) is typically indicative for the presence of structural defects \cite{FerrariNN8_2013,YoonJKPS55_2009}. Nevertheless, this observation is not surprising due to the fact that the graphene  layers are warped around the Co nanorods with an average diameter of only 20~nm.

The resonance observed at $\omega_{\stext{2D}}=2697$$\pm$1~cm$^{-1}$ is attributed to the second order D-band contribution. It is well known that the Raman 2D-band is particularly sensitive to the number of graphene layers and is furthermore closely correlated with the electronic band structure of the material \cite{FerrariSSC143_2007,YoonJKPS55_2009,FerrariNN8_2013}. For single layer graphene the the 2D resonance is located at $\omega_{\stext{2D}}$=2685$\pm$1~cm$^{-1}$. With increasing number of graphene layers the 2D resonance shifts to higher energies and can be observed for graphite as a composition of two contributions at $\omega_{\stext{2D,1}}\approx2720$~cm$^{-1}$ and  $\omega_{\stext{2D,2}}\approx2740$~cm$^{-1}$ \cite{VidanoSSC39_1981}. Based on the 2D peak position, the graphene grown on the slanted Co nanocolumns is likely to be 2-3 layers thick. A Bernal-stacked graphene of such thickness exhibits much higher $I_{\stext{2D}}/I_{\stext{G}}$ ratios than observed in Fig.~\ref{fig:raman},\cite{FerrariNN8_2013} but graphene layers grown on nanoscopic columns are likely twisted, and the $I_{\stext{2D}}/I_{\stext{G}}$ ratio in bilayer graphene was recently shown to strongly depend the twisting angle \cite{KimPRL108_2012, HavenerNL12_2012, HeNL13_2013}. The presence of the structural defects, which is consistent with the high intensity of the D band, could also contribute to the lowered $I_{\stext{2D}}/I_{\stext{G}}$ ratio \cite{ChildresNJP13_2011}. We therefore conclude that the CVD process resulted in few-layer graphene with an estimated number of monolayers (ML) between 2 and 3 and a substantial structural disorder, which is attributed to the 3D nanostructured surface.

Angle-resolved GSE data determine structural and anisotropic optical properties of the Co SCTF before and after the CVD process. It has been demonstrated that the permittivity tensor of highly spatially coherent slanted nanocolumns can be accurately described from THz frequencies to the ultra-violet using an anisotropic Bruggeman effective medium approximation (AB-EMA)  \cite{HofmannAPL99_2011,SchmidtAPL100_2012,SchmidtJAP114_2013}.
In order to correctly render the optical response of the SCTFs discussed here, the Bruggeman approximation, which was developed for disordered inhomogeneous media with spherical inclusions \cite{BruggemanAP24_1935}, is augmented to account for highly spatially coherent, oriented elliptical inclusions \cite{AspnesTSF89_1982,SmithOC71_1989,Sihvola_1999, SchmidtAPL94_2009,SchmidtAPL100_2012}.
The effective dielectric function tensor described by the AB-EMA is composed of three major components $\varepsilon_{\stext{eff}, a}$, $\varepsilon_{\stext{eff}, b}$, and $\varepsilon_{\stext{eff}, c}$  along the major axes $\bm{a}, \bm{b},$ and $\bm{c}$ of an orthorhombic system. $\varepsilon_{\stext{eff}, j}$ with $j=a,b,$ and $c$ are given in implicit form by:\cite{BergnerJOSAA27_2010,SchmidtAPL100_2012}
\begin{equation}
\label{eq:AB-EMA}
  \sum_{n=1}^m{f_n\frac{\veps_{n}-\veps_{\stext{eff},j}}{\veps_{\stext{eff},j}+L_j(\veps_{n}-\veps_{\stext{eff},j})}}=0.
\end{equation}
In Eqn.~(\ref{eq:AB-EMA}) the material constituents' dielectric permittivity and volume fraction are denoted by $\veps_{n}$ and $f_n$, respectively. For the as-grown Co SCTF the corresponding AB-EMA consists of two material contributions: the permittivity of the host medium (air) $\veps_{\stext{1}}=1$ and the permittivity of the nanorods (cobalt) $\veps_{\stext{2}}$. In the case of the post-CVD SCTFs the graphene permittivity is included in the AB-EMA as a third component $\veps_{\stext{3}}$. The factors $L_j$ render the  depolarization of the elliptical inclusions along the major polarizability axes $\bm{a},\bm{b},\bm{c}$ independent of the ellipsoid shape the sum-rule where $L_a+L_b+L_c=1$ must be satisfied \cite{Sihvola_1999}.
The collective response of slanted nanocolumnar arrays may exhibit quasi monoclinic properties \cite{SchmidtJAP105_2009,SchmidtOL34_2009,SchmidtAPL94_2009,SchmidtJAP114_2013}. Such an effect is described through projection of the orthogonal basis system onto the monoclinic in which the semiaxis $\bm{b}$ is tilted towards $\bm{c}$ by a monoclinic angle $\beta$ as described in Ref.~\onlinecite{SchmidtJAP105_2009,SchmidtAPL100_2012}.

\begin{table}
\caption{\label{tab:results}
Summary of the best-model parameters obtained from the analysis of the angle-resolved GSE data obtained before (Fig.~\ref{fig:MM} (left)) and after graphene CVD (Fig.~\ref{fig:MM} (right)). The uncertainty of the last digit (90\% reliability) is given in parentheses.\vspace{-0.3cm}}
\begin{tabular}{llp{0cm}cp{0cm}c}
\hline\hline
\vspace{-0.2cm} &&&&&\\
\multicolumn{2}{c}{Parameter}&&as-grown && post-CVD\\
\hline
\vspace{-0.2cm} &&&&&\\
thickness&$d$ (nm)  						&& 90.78(3) 	&& 132.53(7)\\
slanting angle&$\theta$ ($^{\circ}$)  		&& 56.4(1)  	&& 48.06(2)\\
\multirow{2}{*}{volume fraction} &\multicolumn{1}{|l}{$f_{\stext{void}}$(\%)}		&& 76.72(1)  	&& 70.94(2)\\
								 &\multicolumn{1}{|l}{$f_{\stext{MLG}}$(\%)}	&& N/A			&& 5.02(8)\\
                                 monoclinic angle&$\beta$ ($^{\circ}$)		&& 81.06(1)		&& 86.79(1)\\
\multirow{3}{*}{depolarization}&\multicolumn{1}{|l}{$L_{\stext{a}}$}				&& 0.3769(1)	&& 0.3906(1)\\
&\multicolumn{1}{|l}{$L_{\stext{b}}$}				&& 0.4646(1)	&& 0.4036(1)\\
&\multicolumn{1}{|l}{$L_{\stext{c}}$}				&& 0.1585(1)	&& 0.2058(2)\\

\hline\hline
\end{tabular}
\end{table}

\begin{figure*}
\fcapside[12cm]
{\caption{Best-model effective optical constants (refractive index (a); extinction coefficient (b)) along major axes of polarizability $a$, $b$, $c$, for the as-grown Co SCTF (solid lines) in comparison to the post graphene CVD sample (dashed lines). Note that the graphene coating renders the optical response of the sample uniaxial, i.e., the difference in the optical constants along $a$ and $b$ direction vanishes after the graphene growth.}\label{fig:nk}}
{\includegraphics[keepaspectratio=true,width=5.8cm, clip, trim=0 0 0 0 ]{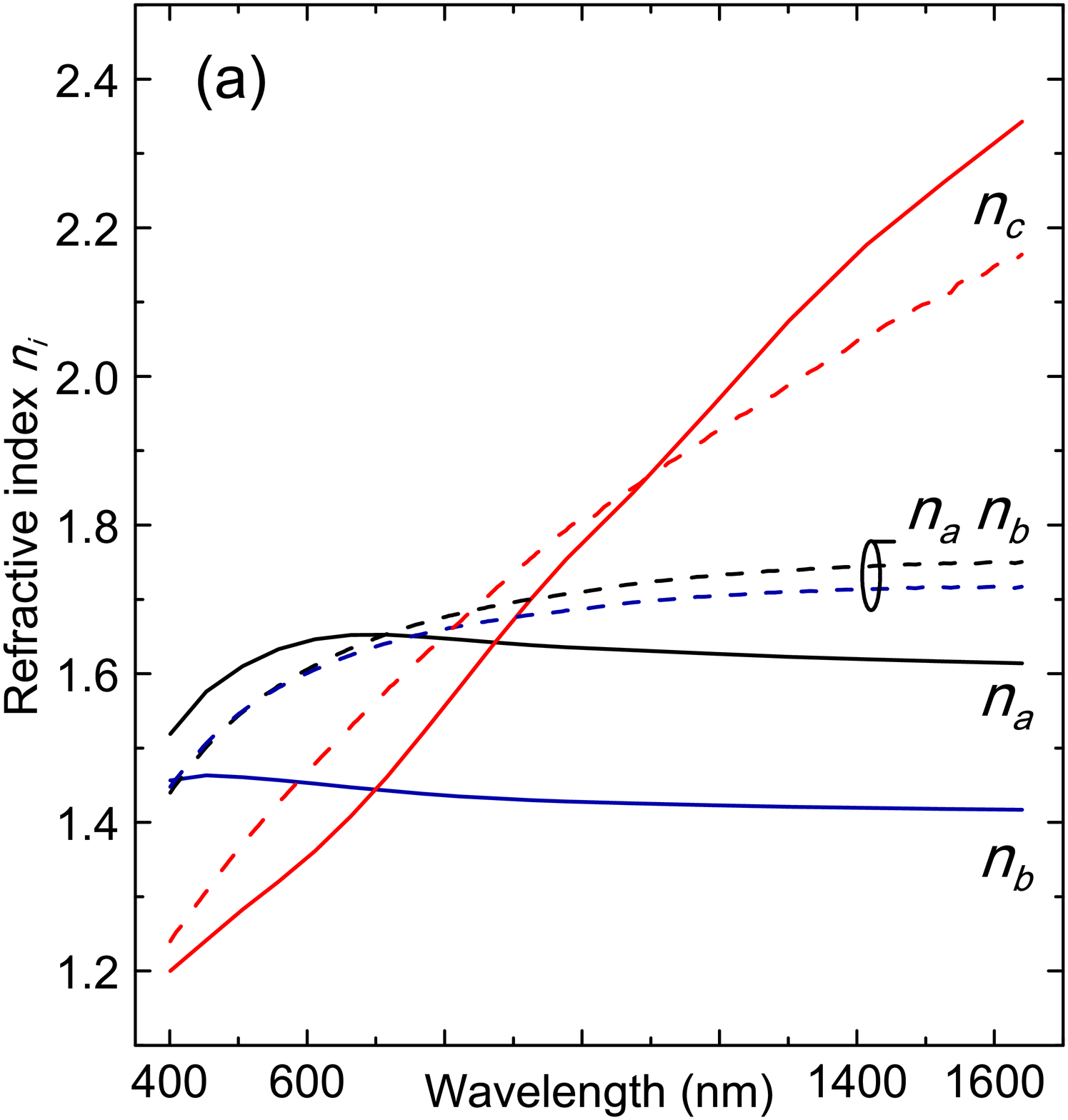}\hspace{0.2cm}\includegraphics[keepaspectratio=true,width=5.9cm, clip, trim=0 0 0 0 ]{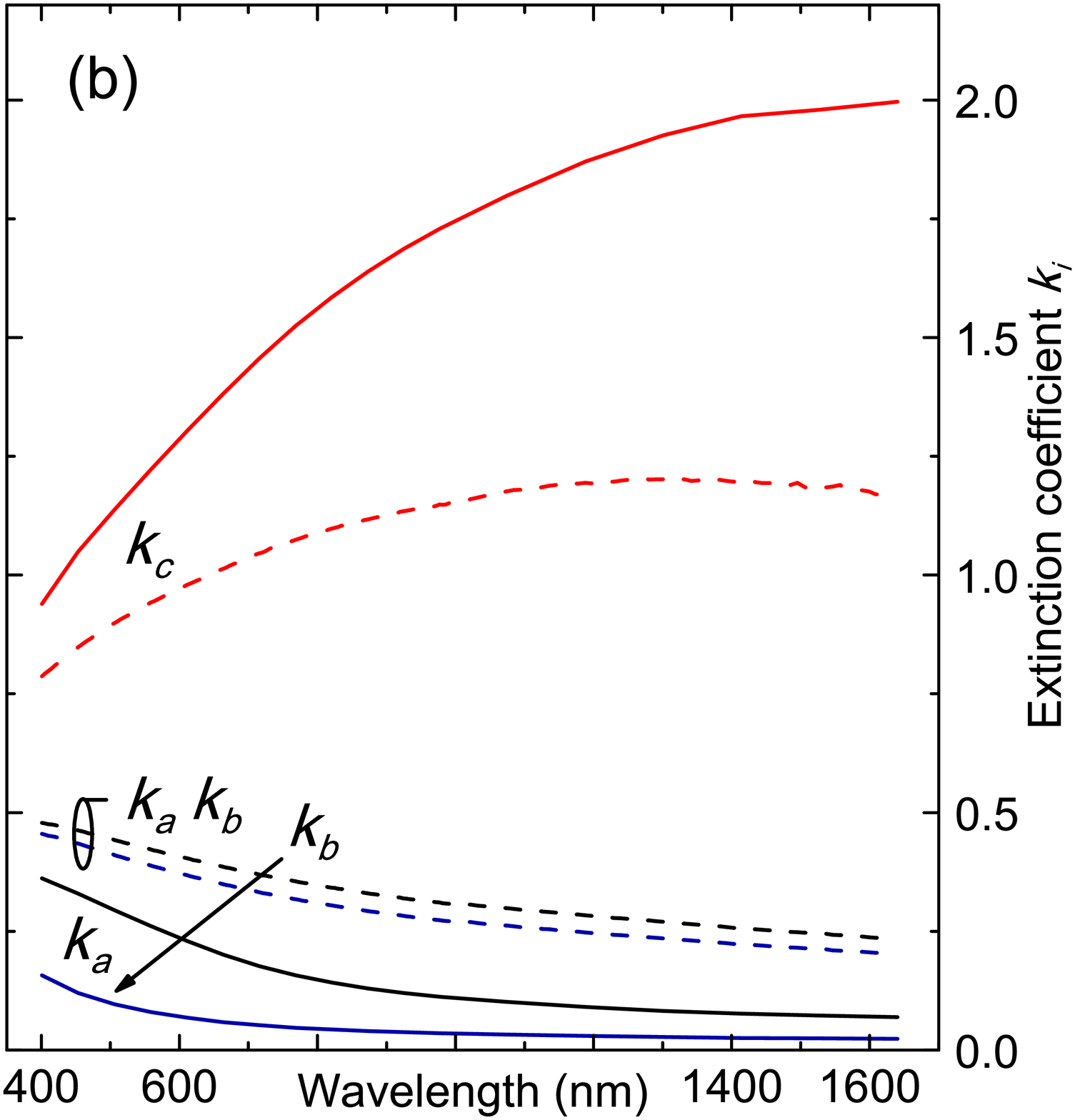}}\vspace{-0.3cm}
\end{figure*}

Figure~\ref{fig:MM} depicts the experimental (symbols) and best-model calculated (solid lines) Mueller matrix data obtained for the Co SCTF before (Fig.~\ref{fig:MM} left) and after graphene deposition (Fig.~\ref{fig:MM} right) as a function of the sample in-plane orientation $\varphi$.
The angle-resolved GSE data sets for each sample were combined within a common model data analysis during which model parameters (see Tab.~\ref{tab:results}) were varied using least-squares approaches, which minimize a weighted test function until calculated and measured data match as close as possible (best-model). A good agreement between the experimental and best-model calculated Mueller matrix data is found using a three-phase stratified layer optical model composed of Si substrate, native SiO$_2$ layer, and AB-EMA based layer for the SCTF. The best-model parameters obtained for both samples are summarized in Tab.~\ref{tab:results}. Literature values were used for the dielectric function of Si substrate, SiO$_2$, and CVD graphene and which were not further varied during the analysis  \cite{HerzingerJAP83_1998,NelsonAPL97_2010,BoosalisAPL101_2012}.
The dielectric function of the cobalt nanorods was varied during the analysis wavelength-by-wavelength to account for differences between bulk cobalt permittivity and GLAD grown material \cite{SchmidtMRSSP2012, SchmidtAPL100_2012}. The obtained results are comparable to those reported earlier  \cite{SchmidtJAP105_2009}.
Comparing the GSE spectra obtained before and after the graphene CVD it can be observed that in particular the so called off-diagonal block Mueller matrix elements ($M_{13}$,$M_{14}$,$M_{23}$,$M_{24}$) are subject to change whereas the on-diagonal Mueller matrix ($M_{12}$,$M_{22}$,$M_{33}$,$M_{34}$) components only scale slightly in their amplitude. This observation is also reflected in the effective optical constants of the SCTF before and after graphene CVD which are depicted in Fig.~\ref{fig:nk}. The conformal graphene coating introduces the strongest changes on the polarizability in $\bm{a}$ and $\bm{b}$ direction while the $\bm{c}$ direction is only lightly shifted. While the as-grown SCTF exhibits biaxial optical properties the graphene coating changes the optical properties of the SCTF to uniaxial.

A very good agreement can be found for the structural parameters obtained by  SEM, Raman, and GSE experiments. In particular, slanting angle and SCTF thickness agree very well between SEM and GSE best-model parameters (Tab.~\ref{tab:results}). Assuming ideal cylindric nanocolumns with a diameter of 20~nm (Fig.~\ref{fig:SEM}) the AB-EMA graphene fraction ($f_{\rm MLG}$) could correspond to a conformal coating with a thickness of 0.9~nm (2.7 ML), which would be in excellent agreement with the 2 to 3 ML estimated from the Raman measurements. We therefore conclude that the graphene coating was successful and GSE and Raman data analysis suggest that this multilayer graphene coating covers the SCTF conformally.

In summary, it has been demonstrated that metal SCTFs can be conformally coated with graphene using a low temperature CVD process and acetylene as a carbon supplying precursor gas. The low temperature CVD process has been found to preserve the nanostructure integrity. The structural parameters obtained by
GSE, Raman spectroscopy and SEM are in very good agreement. In particular, it has been shown that the anisotropic optical response can be described using the AB-EMA approach where structural parameters of graphene-coated cobalt STFs could be extracted. The methodology developed in this work will be useful for the synthesis and characterization of other graphene-coated 3D metallic nanostructures.

The authors would like to acknowledge financial support from the Army Research Office (W911NF-09-C-0097), the National Science Foundation (MRSEC DMR-0820521, MRI-CMMI 1337856, DMR-0907475, ECCS-0846329, EPS-1004094), the Nebraska Center for Energy Sciences Research (No.~12-00-13), the Nebraska Research Initiative, the Nebraska Center for Materials and Nanoscience, and the J.A.~Woollam Foundation.


\begin{thebibliography}{10}
\providecommand{\bibnamefont}[1]{#1}
\providecommand{\bibfnamefont}[1]{#1}

\bibitem{RodenhausenOE20_2012}
\bibfnamefont{K.~B.} \bibnamefont{Rodenhausen},
  \bibfnamefont{D.}~\bibnamefont{Schmidt},
  \bibfnamefont{T.}~\bibnamefont{Kasputis}, \bibfnamefont{A.~K.}
  \bibnamefont{Pannier}, \bibfnamefont{E.}~\bibnamefont{Schubert},
  \bibnamefont{and} \bibfnamefont{M.}~\bibnamefont{Schubert}, Opt. Express
  \textbf{20}, 5419 (2012).

\bibitem{KasputisJCC_2013}
\bibfnamefont{T.}~\bibnamefont{Kasputis},
  \bibfnamefont{M.}~\bibnamefont{Koenig},
  \bibfnamefont{D.}~\bibnamefont{Schmidt},
  \bibfnamefont{D.}~\bibnamefont{Sekora}, \bibfnamefont{K.~B.}
  \bibnamefont{Rodenhausen}, \bibfnamefont{K.-J.} \bibnamefont{Eichhorn},
  \bibfnamefont{P.}~\bibnamefont{Uhlmann},
  \bibfnamefont{E.}~\bibnamefont{Schubert}, \bibfnamefont{A.~K.}
  \bibnamefont{Pannier}, \bibfnamefont{M.~M.} \bibnamefont{Schubert},
  \bibnamefont{and} \bibfnamefont{M.}~\bibnamefont{Stamm}, J. Phys. Chem. C
  \textbf{117}, 13971 (2013).

\bibitem{LeeS321_2008}
\bibfnamefont{C.}~\bibnamefont{Lee}, \bibfnamefont{X.}~\bibnamefont{Wei},
  \bibfnamefont{J.~W.} \bibnamefont{Kysar}, \bibnamefont{and}
  \bibfnamefont{J.}~\bibnamefont{Hone}, Science \textbf{321}, 385 (2008).

\bibitem{BunchS315_2007}
\bibfnamefont{J.~S.} \bibnamefont{Bunch}, \bibfnamefont{A.~M.}
  \bibnamefont{van~der Zande}, \bibfnamefont{S.~S.} \bibnamefont{Verbridge},
  \bibfnamefont{I.~W.} \bibnamefont{Frank}, \bibfnamefont{D.~M.}
  \bibnamefont{Tanenbaum}, \bibfnamefont{J.~M.} \bibnamefont{Parpia},
  \bibfnamefont{H.~G.} \bibnamefont{Craighead}, \bibnamefont{and}
  \bibfnamefont{P.~L.} \bibnamefont{McEuen}, Science \textbf{315}, 490 (2007).

\bibitem{RuoffACSN5_2011}
\bibfnamefont{S.}~\bibnamefont{Chen}, \bibfnamefont{L.}~\bibnamefont{Brown},
  \bibfnamefont{M.}~\bibnamefont{Levendorf},
  \bibfnamefont{W.}~\bibnamefont{Cai}, \bibfnamefont{S.-Y.} \bibnamefont{Ju},
  \bibfnamefont{J.}~\bibnamefont{Edgeworth},
  \bibfnamefont{X.}~\bibnamefont{Li}, \bibfnamefont{C.~W.}
  \bibnamefont{Magnuson}, \bibfnamefont{A.}~\bibnamefont{Velamakanni},
  \bibfnamefont{R.~D.} \bibnamefont{Piner},
  \bibfnamefont{J.}~\bibnamefont{Kang}, \bibfnamefont{J.}~\bibnamefont{Park},
  \bibfnamefont{and} \bibfnamefont{R.~S.}~\bibnamefont{Ruoff}, ACS Nano \textbf{5}, 1321 (2011).

\bibitem{BolotinACSN6_2012}
\bibfnamefont{D.}~\bibnamefont{Prasai}, \bibfnamefont{J.~C.}
  \bibnamefont{Tuberquia}, \bibfnamefont{R.~R.} \bibnamefont{Harl},
  \bibfnamefont{G.~K.} \bibnamefont{Jennings}, \bibnamefont{and}
  \bibfnamefont{K.~I.} \bibnamefont{Bolotin}, ACS Nano \textbf{6}, 1102 (2012).

\bibitem{FarmerNL9_2009}
\bibfnamefont{D.~B.} \bibnamefont{Farmer},
  \bibfnamefont{R.}~\bibnamefont{Golizadeh-Mojarad},
  \bibfnamefont{V.}~\bibnamefont{Perebeinos}, \bibfnamefont{Y.-M.}
  \bibnamefont{Lin}, \bibfnamefont{G.~S.} \bibnamefont{Tulevski},
  \bibfnamefont{J.~C.} \bibnamefont{Tsang}, \bibnamefont{and}
  \bibfnamefont{P.}~\bibnamefont{Avouris}, Nano Lett. \textbf{9}, 388 (2009).

\bibitem{SinitskiiACSN4_2010}
\bibfnamefont{A.}~\bibnamefont{Sinitskii},
  \bibfnamefont{A.}~\bibnamefont{Dimiev}, \bibfnamefont{D.~A.}
  \bibnamefont{Corley}, \bibfnamefont{A.~A.} \bibnamefont{Fursina},
  \bibfnamefont{D.~V.} \bibnamefont{Kosynkin}, \bibnamefont{and}
  \bibfnamefont{J.~M.} \bibnamefont{Tour}, ACS Nano \textbf{4}, 1949 (2010).

\bibitem{SchmidtOL34_2009}
\bibfnamefont{D.}~\bibnamefont{Schmidt}, \bibfnamefont{B.}~\bibnamefont{Booso},
  \bibfnamefont{T.}~\bibnamefont{Hofmann},
  \bibfnamefont{E.}~\bibnamefont{Schubert},
  \bibfnamefont{A.}~\bibnamefont{Sarangan}, \bibnamefont{and}
  \bibfnamefont{M.}~\bibnamefont{Schubert}, Opt. Lett. \textbf{34}, 992 (2009).

\bibitem{SchmidtJAP105_2009}
\bibfnamefont{D.}~\bibnamefont{Schmidt}, \bibfnamefont{A.~C.}
  \bibnamefont{Kjerstad}, \bibfnamefont{T.}~\bibnamefont{Hofmann},
  \bibfnamefont{R.}~\bibnamefont{Skomski},
  \bibfnamefont{E.}~\bibnamefont{Schubert}, \bibnamefont{and}
  \bibfnamefont{M.}~\bibnamefont{Schubert}, J. Appl. Phys. \textbf{105}, 113508
  (2009).

\bibitem{SchmidtAPL94_2009}
\bibfnamefont{D.}~\bibnamefont{Schmidt}, \bibfnamefont{B.}~\bibnamefont{Booso},
  \bibfnamefont{T.}~\bibnamefont{Hofmann},
  \bibfnamefont{E.}~\bibnamefont{Schubert},
  \bibfnamefont{A.}~\bibnamefont{Sarangan}, \bibnamefont{and}
  \bibfnamefont{M.}~\bibnamefont{Schubert}, Appl. Phys. Lett. \textbf{94},
  011914 (2009).

\bibitem{SchmidtAPL100_2012}
\bibfnamefont{D.}~\bibnamefont{Schmidt},
  \bibfnamefont{E.}~\bibnamefont{Schubert}, \bibnamefont{and}
  \bibfnamefont{M.}~\bibnamefont{Schubert}, Appl. Phys. Lett. \textbf{100},
  011912 (2012).

\bibitem{SchmidtJAP114_2013}
\bibfnamefont{D.}~\bibnamefont{Schmidt} \bibnamefont{and}
  \bibfnamefont{M.}~\bibnamefont{Schubert}, J. Appl. Phys. \textbf{114}, 083510
  (2013).

\bibitem{WilsonJMCC2013}
\bibfnamefont{P.~M.} \bibnamefont{Wilson},
  \bibfnamefont{G.}~\bibnamefont{Mbah}, \bibfnamefont{T.~G.}
  \bibnamefont{Smith}, \bibfnamefont{D.}~\bibnamefont{Schmidt},
  \bibfnamefont{M.}~\bibnamefont{Shekhirev}, \bibfnamefont{R.~Y.}
  \bibnamefont{Lai}, \bibfnamefont{T.}~\bibnamefont{Hofmann}, \bibnamefont{and}
  \bibfnamefont{A.}~\bibnamefont{Sinitskii}, J. Mat. Chem. C, \textit{accepted}, \texttt{DOI:10.1039/C3TC32277G}
  (2013).

\bibitem{FerrariNN8_2013}
\bibfnamefont{A.~C.} \bibnamefont{Ferrari} \bibnamefont{and}
  \bibfnamefont{D.~M.} \bibnamefont{Basko}, Nature Nano. \textbf{8}, 235
  (2013).

\bibitem{FerrariPRL97_2006}
\bibfnamefont{A.~C.}~\bibnamefont{Ferrari}, \bibfnamefont{J.~C.}~\bibnamefont{Meyer},
  \bibfnamefont{V.}~\bibnamefont{Scardaci},
  \bibfnamefont{C.}~\bibnamefont{Casiraghi},
  \bibfnamefont{M.}~\bibnamefont{Lazzeri},
  \bibfnamefont{F.}~\bibnamefont{Mauri},
  \bibfnamefont{S.}~\bibnamefont{Piscanec},
  \bibfnamefont{D.}~\bibnamefont{Jiang},
  \bibfnamefont{K.~S.}~\bibnamefont{Novoselov},
  \bibfnamefont{S.}~\bibnamefont{Roth}, \bibnamefont{and}
  \bibfnamefont{A.~K.}~\bibnamefont{Geim}, Phys. Rev. Lett. \textbf{97}, 187401
  (2006).

\bibitem{FerrariSSC143_2007}
\bibfnamefont{A.~C.} \bibnamefont{Ferrari}, Solid State Commun. \textbf{143},
  47 (2007).

\bibitem{YoonJKPS55_2009}
\bibfnamefont{D.}~\bibnamefont{Yoon}, \bibfnamefont{H.}~\bibnamefont{Moon},
  \bibfnamefont{H.}~\bibnamefont{Cheong}, \bibfnamefont{J.~S.}
  \bibnamefont{Choi}, \bibfnamefont{J.~A.} \bibnamefont{Choi},
  \bibnamefont{and} \bibfnamefont{B.~H.} \bibnamefont{Park}, J. Korean Phys.
  Soc \textbf{55}, 1299 (2009).

\bibitem{VidanoSSC39_1981}
\bibfnamefont{R.~P.} \bibnamefont{Vidano}, \bibfnamefont{D.~B.}
  \bibnamefont{Fischbach}, \bibfnamefont{L.~J.} \bibnamefont{Willis},
  \bibnamefont{and} \bibfnamefont{T.~M.} \bibnamefont{Loehr}, Solid State
  Commun. \textbf{39}, 341  (1981).

\bibitem{KimPRL108_2012}
\bibfnamefont{K.}~\bibnamefont{Kim}, \bibfnamefont{S.}~\bibnamefont{Coh},
  \bibfnamefont{L.~Z.} \bibnamefont{Tan},
  \bibfnamefont{W.}~\bibnamefont{Regan}, \bibfnamefont{J.~M.}
  \bibnamefont{Yuk}, \bibfnamefont{E.}~\bibnamefont{Chatterjee},
  \bibfnamefont{M.}~\bibnamefont{Crommie}, \bibfnamefont{M.~L.}
  \bibnamefont{Cohen}, \bibfnamefont{S.~G.} \bibnamefont{Louie},
  \bibnamefont{and} \bibfnamefont{A.}~\bibnamefont{Zettl}, Phys. Rev. Lett.
  \textbf{108}, 246103 (2012).

\bibitem{HavenerNL12_2012}
\bibfnamefont{R.~W.} \bibnamefont{Havener},
  \bibfnamefont{H.}~\bibnamefont{Zhuang},
  \bibfnamefont{L.}~\bibnamefont{Brown}, \bibfnamefont{R.~G.}
  \bibnamefont{Hennig}, \bibnamefont{and} \bibfnamefont{J.}~\bibnamefont{Park},
  Nano Lett. \textbf{12}, 3162 (2012).

\bibitem{HeNL13_2013}
\bibfnamefont{R.}~\bibnamefont{He}, \bibfnamefont{T.-F.} \bibnamefont{Chung},
  \bibfnamefont{C.}~\bibnamefont{Delaney},
  \bibfnamefont{C.}~\bibnamefont{Keiser}, \bibfnamefont{L.~A.}
  \bibnamefont{Jauregui}, \bibfnamefont{P.~M.} \bibnamefont{Shand},
  \bibfnamefont{C.}~\bibnamefont{Chancey},
  \bibfnamefont{Y.}~\bibnamefont{Wang}, \bibfnamefont{J.}~\bibnamefont{Bao},
  \bibnamefont{and} \bibfnamefont{Y.~P.} \bibnamefont{Chen}, Nano Lett.
  \textbf{13}, 3594 (2013).

\bibitem{ChildresNJP13_2011}
\bibfnamefont{I.}~\bibnamefont{Childres}, \bibfnamefont{L.~A.}
  \bibnamefont{Jauregui}, \bibfnamefont{J.}~\bibnamefont{Tian},
  \bibnamefont{and} \bibfnamefont{Y.~P.} \bibnamefont{Chen}, New J. Phys.
  \textbf{13}, 025008 (2011).

\bibitem{HofmannAPL99_2011}
\bibfnamefont{T.}~\bibnamefont{Hofmann},
  \bibfnamefont{D.}~\bibnamefont{Schmidt},
  \bibfnamefont{A.}~\bibnamefont{Boosalis},
  \bibfnamefont{P.}~\bibnamefont{K\"{u}hne},
  \bibfnamefont{R.}~\bibnamefont{Skomski}, \bibfnamefont{C.~M.}
  \bibnamefont{Herzinger}, \bibfnamefont{J.~A.} \bibnamefont{Woollam},
  \bibfnamefont{M.}~\bibnamefont{Schubert}, \bibnamefont{and}
  \bibfnamefont{E.}~\bibnamefont{Schubert}, Appl. Phys. Lett. \textbf{99},
  081903 (2011).

\bibitem{BruggemanAP24_1935}
\bibfnamefont{D.}~\bibnamefont{Bruggeman}, Ann. Phys. \textbf{416}, 636 (1935).

\bibitem{AspnesTSF89_1982}
\bibfnamefont{D.}~\bibnamefont{Aspnes}, Thin Solid Films \textbf{89}, 249
  (1982).

\bibitem{SmithOC71_1989}
\bibfnamefont{G.}~\bibnamefont{Smith}, Opt. Commun. \textbf{71}, 279  (1989).

\bibitem{Sihvola_1999}
\bibfnamefont{A.}~\bibnamefont{Sihvola}, \textit{Electromagnetic mixing formulas
  and applications} (The Institution of Electrical Engineers, London, 1999).

\bibitem{BergnerJOSAA27_2010}
\bibfnamefont{B.~C.} \bibnamefont{Bergner}, \bibfnamefont{T.~A.}
  \bibnamefont{Germer}, \bibnamefont{and} \bibfnamefont{T.~J.}
  \bibnamefont{Suleski}, J. Opt. Soc. Am. A \textbf{27}, 1083 (2010).

\bibitem{HerzingerJAP83_1998}
\bibfnamefont{C.~M.} \bibnamefont{Herzinger},
  \bibfnamefont{B.}~\bibnamefont{Johs}, \bibfnamefont{W.~A.}
  \bibnamefont{McGahan}, \bibnamefont{and} \bibfnamefont{a.~W.~P.}
  \bibnamefont{J.~A.~Woollam}, J. Appl. Phys. \textbf{83}, 3323 (1998).

\bibitem{NelsonAPL97_2010}
\bibfnamefont{F.~J.} \bibnamefont{Nelson}, \bibfnamefont{V.~K.}
  \bibnamefont{Kamineni}, \bibfnamefont{T.}~\bibnamefont{Zhang},
  \bibfnamefont{E.~S.} \bibnamefont{Comfort}, \bibfnamefont{J.~U.}
  \bibnamefont{Lee}, \bibnamefont{and} \bibfnamefont{A.~C.}
  \bibnamefont{Diebold}, Appl. Phys. Lett. \textbf{97}, 253110 (2010).

\bibitem{BoosalisAPL101_2012}
\bibfnamefont{A.}~\bibnamefont{Boosalis},
  \bibfnamefont{T.}~\bibnamefont{Hofmann},
  \bibfnamefont{V.}~\bibnamefont{Darakchieva},
  \bibfnamefont{R.}~\bibnamefont{Yakimova}, \bibnamefont{and}
  \bibfnamefont{M.}~\bibnamefont{Schubert}, Appl. Phys. Lett. \textbf{101},
  011912 (2012).

\bibitem{SchmidtMRSSP2012}
\bibfnamefont{D.}~\bibnamefont{Schmidt},
  \bibfnamefont{E.}~\bibnamefont{Schubert}, \bibnamefont{and}
  \bibfnamefont{M.}~\bibnamefont{Schubert}, Mat. Res. Soc. Symp. Proc.
  \textbf{1408}, CC13 (2012).

\end{thebibliography}

\end{document}